\documentstyle[psfig,epsfig,aps,multicol,fancyheadings]{revtex} 
\begin{document}
\title{Minimum spanning trees on random networks}
\author{R. Dobrin \footnote{ \noindent 
{\bf email:} dobrin@pa.msu.edu} and P.M. Duxbury  }

\address{ Dept. of Physics/Ast. and 
Center for Fundamental Materials Research,\\
Michigan State University, East Lansing, MI 48824, USA.}
\maketitle

\begin{abstract}
We show that the geometry of minimum spanning trees 
(MST) on random graphs is universal.
Due to this geometric universality, 
we are able to characterise the energy of MST using a
scaling distribution ($P(\epsilon)$) found using uniform disorder.
We show that the MST energy for other disorder distributions is simply related to $P(\epsilon)$.
We discuss the relationship to invasion percolation (IP), to the
directed polymer in a random media (DPRM) and
the implications for the broader issue of universality in disordered systems.

\end{abstract}


\begin{multicols}{2}[]
\narrowtext

Universality is an unproven hypothesis in
disordered systems, though it has been
widely assumed in the development of theories
for the random-field Ising model and for spin glasses\cite{young}.
The hope has been that scaling exponents in
disordered systems should not depend on the
nature of the disorder, provided it is uncorrelated
and the disorder distribution is not too broad.
Percolation\cite{stauffer} and the directed polymer in a random medium
(DPRM)\cite{halpin} reassure us that universality does hold.  However,
the random field Ising model(RFIM) has recently provided
a counter example\cite{swift,angles,duxbury}. In particular we showed that
the mean-field RFIM is non-universal in the ground state as 
the order parameter exponent can depend
continuously on the details of the disorder\cite{duxbury}.  In 
contrast we show here that the MST is {\it superuniversal}
in the sense that the MST geometry is unaltered
even if the distribution of disorder is made very
broad.  Due to this fact, the energy of a MST
can be found from a universal function, for a given
graph topology.  The MST geometry is important in the
{\it strong disorder limit}\cite{cieplak1,cieplak2},
for example as a model
for spin glasses\cite{newman} and for  
hopping transport at low temperatures\cite{cieplak1,tyc,dyre}.
As we shall discuss below, the paths on the minimum spanning tree 
are those on which the {\it energy barrier} is smallest and
for this reason MST paths dominate the kinetics
 at low temperatures\cite{cieplak1,tyc,dyre}.

Perhaps the simplest non-trivial optimisation problem,
the minimum spanning tree(MST) 
is a tree which visits every site in
a graph so that the sum of the costs of the edges in
the tree is minimal(see Fig. 1).  In physics terminology each
edge, $(ij)$ has an energy, $\epsilon_{(ij)}$, and the total energy
is the sum of the energies of the bonds which make up the
minimum spanning tree, ie.
\begin{equation} 
E_{MST}=\sum_{(ij)\  in\  tree} \epsilon_{(ij)}.
\end{equation}  
Due to its practical applications
in a variety of contexts, including image analysis, transportation
networks etc, this problem has been heavily studied by the
engineering community.  This problem is also one
of the most fundamental problems in combinatorial optimisation
and has been intensively studied in the computer science
and applied math communities\cite{cormen}.  The physics community has
been less aware of this problem,
 with notable exceptions\cite{chayes,barabasi,aizenman},
 though it has close
connections to the problem of a fluid invading a
porous medium, as modeled by the invasion percolation(IP)
process.  However 
invasion percolation is a dynamic process which
{\it grows} minimum spanning trees, whereas the
MST itself is a global minimum of a cost function.
The MST must visit {\it every} site in the
graph and so corresponds to continuing
the invasion process until every site in
a finite graph is reached.  This is 
not usually studied in invasion percolation,
in which the steady state regime in a very
large lattice is of most interest\cite{wilkinson,sheppard}.

We concentrate on two aspects of minimal spanning
trees on random graphs: the geometry of the 
{\it paths} on the minimum spanning tree; 
and the {\it cost} of the minimal
spanning tree.  First we show that the geometry
of minimum spanning trees is {\it universal}
 even when the disorder is broad.

To demonstrate the universality of MST,
it is useful to first describe how MST's are found
in practice. For simplicity, consider 
square and cubic lattices whose edges are assigned
costs (energies) drawn from a uniform distribution on 
the interval $[0,1]$.  In order to find the minimal
spanning tree on such graphs, we use Prim's algorithm
which is a greedy algorithm (in physics these are called
growth, invasion  or extremal algorithms) which chooses the
best site for advance at each timestep.  In the
computer science literature\cite{cormen} Prim's method starts
by choosing the cheapest bond in the whole graph, 
and then by growing outward to the cheapest bond 
which is adjacent to the starting bond.  Each
bond which is invaded is added to the growing cluster
and the process is iterated until every site has 
been reached.  Bonds can only be invaded if they 
do not produce a cycle, so that the tree structure is
maintained.  However it
is not essential to start from the cheapest bond
as growth starting from any site leads to the
same MST.   This latter process is
identical to {\it bond invasion percolation}, which
then finds the MST exactly.
Intuitively, invasion algorithm finds the exact minimum 
spanning tree because each site in a MST must be visited
at least once and the IP algorithm selects 
the best way to make this choice at each site. 
 It is a standard
exercise in algorithm theory 
to prove this rigorously\cite{cormen}.

The {\it property which makes the 
MST geometry universal} is that
to produce a unique MST in a graph
we need only specify  {\it a unique ordering of
the bonds of the graph, according to their energies}.
 It does not matter if the bond energies are
nearly the same or wildly different,
it is only the ordering of the energies that matters. 
This can be intuitively understood by
making a list of the bonds
ordered from the smallest in energy
to the largest.  Now sequentially remove
bonds, starting with the largest in energy,
however with the
rule that the removal of a bond {\it cannot
disconnect the graph into two pieces}(this
ensures that we end up with a spanning tree).
Continue this process until no bonds
can be removed.  This final state is the
minimum spanning tree and is very
similar to the algorithm suggested by
Cieplak et al. for this problem\cite{cieplak1}
(the invasion method is much more efficient
though).  All that matters for this bond
removal process is the {\it ordering}
of the bond energies, and hence the
geometry of the final tree so formed
only depends on that ordering.

Given the fact that a MST only
depends on the ordering of the bond 
energies, any transformation 
$\epsilon\rightarrow f(\epsilon)$ 
which preserves the ordering of the bond energies (eg. the
bond which has the fiftieth largest energy is the same before
and after the transformation) leaves the 
MST geometry unaltered.  Now note that if  
$f(\epsilon)$ is any {\it non-decreasing function} of $\epsilon$, the
ordering of a set $\{\epsilon_1...\epsilon_n\}$ is unaltered
under the transformation to $\{f(\epsilon_1)...f(\epsilon_n)\}$.  
This observation is germane to the issue of
universality due to the fact that 
we can use the uniform distribution to  
sample according to a general distribution 
$F(x)$, by assigning $F(x)dx = dy$ which
transforms the interval $dy$ of the uniform distribution
to the interval $F(x)dx$ of the general distribution.
Thus if we randomly choose a number $y$ from the uniform
distribution, the corresponding random number
from the distribution $F(x)$ is 
\begin{equation}
x = G^{-1}(y)\ \ \  {\rm where}\ \ \ G(x)=\int_0^x F(x') dx'.
\end{equation}
Now note that $G(x)$ is a {\it non-decreasing} 
function of $x$ because $F(x)$ is a probability
and so is positive which implies that $G^{-1}(y)$
is also a non-decreasing function of $y$(provided $G$
is invertible). 
Thus the transformation(2) preserves
the {\it ordering of the bond costs} and
hence the geometry of MST 
is the same for {\it any probability
distribution}. This is one of the
few non-trivial problems for which
universality to disorder can be
explicitly demonstrated.

Numerical calculation of the geometry of the
paths on a MST is carried out as illustrated
in Fig. 1.   The number of bonds, $n$,
which lie on a MST path between two sites which are
separated by Euclidean distance $l$, is
found to scale as $l^y$, where 
$y=1.22\pm 0.01$ (square lattice) and
$y=1.42\pm 0.02$ (cubic lattice)(see Fig. 2). 
The strands in invasion
percolation\cite{cieplak2} give exactly the same
scaling.  In fact the paths on MST scale
with the same fractal dimension throughout the
Prim growth process and only the form
of the scaling distribution changes (see Figs. 2a and 2b).
A more detailed study of the form of these scaling
distributions will be discussed elsewhere.

Having proven that the geometry of MST is 
fractal and universal, we now show that 
it is possible to find
the energy of MST from one universal function,
for a given graph topology.
Numerical results for the
appropriate universal function are presented
in Figs. 3a,b 
for minimum spanning trees on 
square and cubic lattices.  The function
we plot is the probability, $P(\epsilon)$ 
that a bond of cost $\epsilon$ for a uniform
distribution(on the interval $[0,1]$),
 lies on the minimum spanning 
tree.  The dashed lines in Fig. 3 are the 
cost distributions for the MST.
 The solid lines are for invasion percolation.  
In invasion percolation $P(\epsilon)$ is 
the {\it acceptance function}
for the case of bond invasion.  In this case, it 
has the interesting property that in the
scaling limit $P(\epsilon) \rightarrow 0$ for $\epsilon>p_c$  
($p_c$ is the bond percolation threshold). 
 However, the MST must reach every site 
 of the graph, in which case it is 
necessary to include bonds which
have $\epsilon>p_c$ (see the dashed lines in Figs. 3a,b).  

From the acceptance probability $P(\epsilon)$, the
total energy of the minimum spanning tree 
with bonds drawn from a uniform distribuiton
is simply (in the scaling limit),
\begin{equation}
E = \int_0^1 \epsilon P(\epsilon)d\epsilon.
\end{equation}
As shown above, the geometry of the minimum
spanning tree is unaltered if we make a 
transformation $\epsilon\rightarrow f(\epsilon)$ of the
bond costs provided $f(\epsilon)$ is {\it a nondecreasing function}
of the bond costs. 
 After making this transformation
 the energy is simply $E = \int_0^1 f(\epsilon) P(\epsilon) d\epsilon$.
In addition, using the arguments given
above we may generalise to the case of an
arbitrary distribution $F(\epsilon)$, in which case(from Eq. (2)) 
\begin{equation}
E_F = \int_0^1 G^{-1}(\epsilon) P(\epsilon) d\epsilon,
\end{equation}
for any disorder distribution $F(x)$.

As stated above, paths on the MST are those
on which the bond of maximum energy is minimal.
In physical terms MST paths are those on which the  
{\it barrier} is minimal.
The barrier on such a path is the {\it largest
cost bond} which lies on that path.  In the 
steady state limit
\begin{equation}
\epsilon_{barrier} \rightarrow p_c,
\end{equation}
that is, the barrier on invasion percolation(IP) paths, in a graph with
edge costs drawn from a uniform distribution, takes on a value
equal to the percolation threshold on that graph.  For other
distributions of disorder, the barrier on IP paths becomes
$\epsilon_{barrier} = G^{-1}(p_c)$ (from Eq.(2)).
However the barrier on {\it typical} MST paths
have $\epsilon_{barrier} > p_c$
for $l\rightarrow \infty$, and it is only on the fractal
IP subset of paths on which Eq. (5) holds.    
Thus hopping transport at low temperatures
will typically occur on IP paths, which 
justifies the use of percolation models in
the calculation of diffusivity and conductivity
in the strong disorder limit\cite{tyc,dyre}.

It is interesting to compare the behavior of paths
on the MST, with the behavior of the directed
polymer in a random medium (DPRM)\cite{halpin}.
The DPRM, and the associated Kardar-Parisi-Zhang growth
process\cite{kardar}, has become a paradigm in the study of disordered
systems.  More recently it has been
noted that DPRM is a subset
of the shortest path(SP) problem in computer science
and engineering\cite{cormen,marsili,dobrin}.  The statement of the
SP problem is deceptively
similar to that of the MST problem discussed above, however
its properties are radically different.   The shortest
path problem seeks to find a path between two
sites in a graph such that the {\it sum of the 
bond costs is minimal}, so that,
\begin{equation} 
E_{SP}=\sum_{(ij)\  in\  path} \epsilon_{(ij)}.
\end{equation}  
If one seeks the shortest
path from a starting {\it source site} to all other sites
in a graph, then a {\it shortest path tree}(SPT)
is formed.  The total cost of the shortest
path tree is the sum of the costs of all of the
paths in the tree.  Note however that in this sum
it is inevitable that some of the bond costs will
appear more than once.   In fact bonds near
the source site will be counted many times.  The SP
problem is in the DPRM universality class except for the
limit of strong disorder when it approaches the MST 
problem\cite{cieplak2,porto}.  The crossover to the strong disorder limit
can be analysed explicitly using the generalised energy,
$\sum \epsilon_{(ij)}^m$ instead of Eq. (6).  In the limit 
$m\rightarrow \infty$ the largest energy dominates and hence
the {\it largest barrier} is all that matters.
It is possible to show, using (2) that $m\rightarrow \infty$ 
corresponds to the strong disorder limit.
The SPT is also distinguished by the
fact that there is a {\it different SPT}
for each starting site in the graph, whereas there
is only {\it one MST} for a finite graph with 
continuous disorder.

In summary, the geometry of minimum spanning trees(MST) 
is universal for all disorder distributions 
because a MST is invariant under the transformation
of its edge costs $\epsilon \rightarrow f(\epsilon)$ where $f$ is
a non-decreasing function of the edge cost $\epsilon$.  This
universality enabled us to find a universal cost function
for the uniform distribution (see Fig. 3) which can be used to
calculate the cost of minimum spanning trees, on the
same graph, for any other distribution (using Eq. (4)).  Paths
on the MST are those with minimal {\it barrier} and
in the steady state IP process this barrier approaches
$p_c$ for a uniform distribution and $G^{-1}(p_c)$ for an
arbitrary distribution(where $G$ is given in Eq. (2)).
The MST geometry underlies physics in the
strong disorder limit, implying that in that
limit there is a strong universality.

This work has been supported by
the DOE under contract DE-FG02-90ER45418.


\newpage
\begin{figure}
\centerline{\epsfig{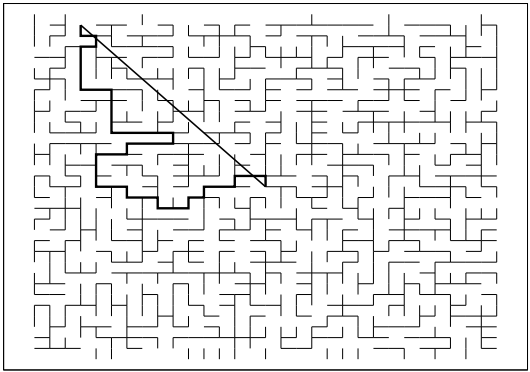}}
\vspace{0.2in}
\caption{The minimum spanning tree(MST) for a 
$30\times 30$ square lattice.
Each edge in the square lattice is assigned an energy 
drawn from the uniform distribution on the
interval [0,1].  Only the bonds on the minimum spanning tree are
drawn in the figure. 
 The wandering heavy line is one path on the MST, starting at the center
of the square lattice. The Euclidean distance between the two ends of this
path is also indicated.}
\label{fig1}
\end{figure}

\begin{figure}
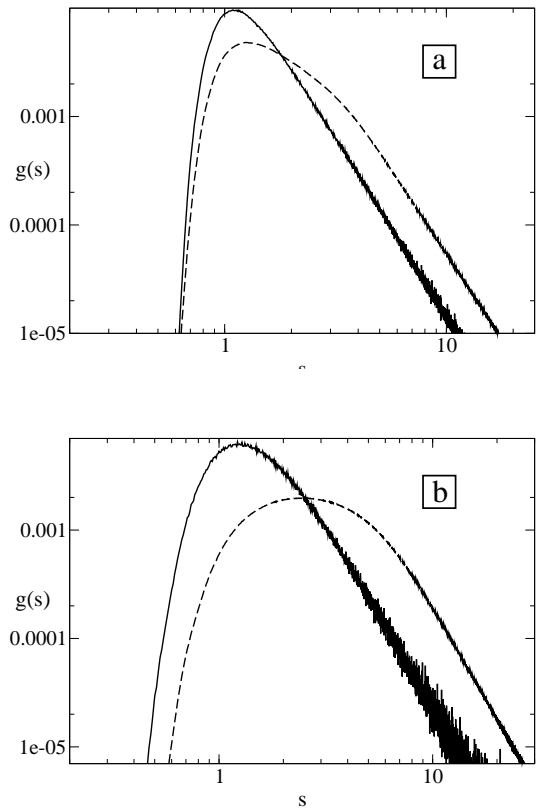

\centerline{\epsfig{file=./2dgs.eps,width=7cm,angle=0}}
\vspace{0.3in}
\centerline{\epsfig{file=./3dgs.eps,width=7cm,angle=0}}
\vspace{0.2in}
\caption{The scaled distributions of pathlengths, $g(s)$, 
on minimum spanning trees on
{\bf a)} square and {\bf b)} cubic lattices.  The scaling variable
is $s=n/l^{D_f}$ where $n$ is the number of bonds in the MST path, $l$ is the
Euclidean distance and $D_f$ is the scaling dimension 
($D_f =1.22\pm 0.01$ (square lattice) 
and $D_f=1.42 \pm 0.02$(cubic lattice)).  The dotted line in these
figures is the scaling distribution on the MST.  For comparison
we also give the scaling distribution for the steady state
during growth of the MST, ie.
 invasion percolation(solid line).  
 In both cases the paths scale
 with the same fractal dimension, in fact this holds at
 all stages of growth of the MST.  The results are
 found from averaging over 2000 realisations of
 $401 \times 401$ square lattices and over 1500
 realisations of $101\times 101 \times 101$ cubic lattices.}
\label{fig2}
\end{figure}

\begin{figure}
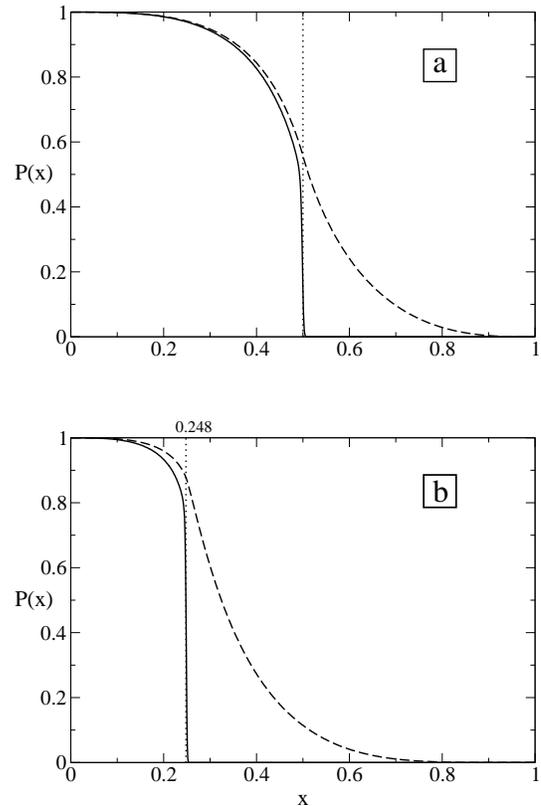

\centerline{\epsfig{file=./2dar.eps,width=7cm,angle=0}}
\vspace{0.2in}
\centerline{\epsfig{file=./3dar.eps,width=7cm,angle=0}}
\vspace{0.2in}
\caption{The probability, $P(x)$, that a bond with energy $x$
lies on the minimum spanning tree for {\bf a)} square and
{\bf b)} cubic lattices.  The case considered here is
the uniform distribution of bond disorder.  The curving dashed lines
are for the MST while the solid lines are for the
steady state during growth of the MST, 
ie. invasion percolation.  The results are
 found from averaging over 1000 realisations of
 $401 \times 401$ square lattices and over 1000
 realisations of $101\times 101 \times 101$ cubic lattices.}
\label{fig3}
\end{figure}
\end{multicols}
\end{document}